\documentclass{article}
\usepackage[utf8]{inputenc}
\usepackage{graphicx}
\usepackage{amsmath,amssymb,bbold,bm,color}
\usepackage{verbatim}
\usepackage{authblk} 

\newcommand{\bee}{\begin{equation}}
\newcommand{\ee}{\end{equation}}

\newcommand{\beq}{\begin{equation}}
\newcommand{\eeq}{\end{equation}}
\newcommand{\nn}{\nonumber \\}
\def\bea{\begin{eqnarray}}
\def\eea{\end{eqnarray}}

\def\sgn{\mathop{\rm sgn}\nolimits}

\title{Simulating black hole quantum dynamics on an optical lattice using the complex Sachdev-Ye-Kitaev model}

\author[1]{Iftekher S. Chowdhury}
\author[2]{Binay Prakash Akhouri}
\author[5]{Shah Haque}
\author[1,3]{Martin H. Bacci}
\author[1,4,5]{Eric Howard}

\affil[1]{Department of Physics and Astronomy, Macquarie University, Sydney, NSW, 2109, Australia}
\affil[2]{Department of Physics, Suraj Singh Memorial College, Ranchi University, Ranchi, Jharkhand, India}
\affil[3]{Maritime and Polytechnic University College, Frederikshavn, Denmark} 
\affil[4]{Swinburne University, Sydney, Australia} 
\affil[5]{Southern Cross Institute, School of Computer Science, Sydney, Australia} 

\date{May 2024}
\begin{document}

\maketitle

\begin{abstract}
We propose a low energy model for simulating an analog black hole on an optical lattice using ultracold atoms. Assuming the validity of the holographic principle, we employ the Sachdev-Ye-Kitaev (SYK) model, which describes a system of randomly infinite range interacting fermions, also conjectured to be an exactly solvable UV-complete model for an extremal black hole in a higher dimensional Anti-de Sitter (AdS) dilaton gravity. At low energies, the SYK model exhibits an emergent conformal symmetry and is dual to the extremal black hole solution in near AdS$_2$ spacetime.  Furthermore, we show how the SYK maximally chaotic behaviour at large $N$ limit, found to be dual to a gauge theory in higher dimensions, can also be employed as a non-trivial investigation tool for the holographic principle. The proposed setup is a theoretical platform to realize the SYK model with relevant exotic effects and behaviour at low energies as a highly non-trivial example of the AdS/CFT duality and a framework for studying black holes.
\end{abstract}

\section{Introduction}

Simulation of quantum matter using ultracold atoms, including fermionic atoms\cite{Giorgini:2008zz} in optical lattices is a very active current area of research, providing a wide range of solutions to unanswered questions in condensed matter physics and beyond\cite{labSYK}.
Over the past ten years, breakthrough experimental work has exposed the dynamics of bosons, that can occupy the same quantum state, and fermions, which cannot occupy the same quantum state, via experiments with ultracold Bose and Fermi gases. 
In particular, integrable (exactly solvable) models, like SYK play a significant role in the understanding of the quantum dynamics in low-dimensional many-body systems\cite{Sachdev-1006}, such as optical lattices with trapped ultracold bosonic and fermonic atoms.
Testing novel concepts and techniques with optical lattices requires development of various geometric structures, superlattices or lattices with disorder that allow comparisons between exactly solvable models and experiment.

The  specific properties of the SYK model have sparked considerable interest in finding an experimental realization of holographic systems. Since no natural materials are known to inherently exhibit the unique interactions of the SYK model, researchers have turned to synthetic quantum systems. A realization of SYK model is a challenging task, which would lead to a better understanding of many-body chaos and its relation to black holes. 
A realistic quantum simulation of the SYK model\cite{Cao:2020rhe} has major stringent conditions and significant bottlenecks, due to the complicated form of perfectly random all-to-all fermionic atom interactions. Several proposals have been put forward on various physical platforms, from topological superconducting wires to graphene flakes in magnetic field. The main challenges include reaching extremely low temperatures in an optical lattice and find an efficient way to extract the information on the quantum many-body state from the experimental data. 
Various experiments using ultracold atoms trapped in optical lattices have succeeded to realize several theoretical models\cite{Wiese:2013uua}, such as Bose-Hubbard model, Lieb-Liniger model or the topological Haldane model. The long term goal is to physically implement the SYK Hamiltonian with cold atom optics that can be used to characterize a non-dispersive system with strong disorder and a flat band spectrum. Here, in particular, the solvable SYK Hamiltonian emerges from an optical Kagome lattice of spinless fermionic atoms with strong disorder. 

The frustration-related physics and the presence of exotic phases exibited by ultracold degenerate Fermi gases in optical lattices are key advantages for simulating SYK toy model. More specifically, we propose a convenient model to simulate near-AdS black hole properties by randomly trapping ultracold atoms in a two-dimensional Kagome optical lattice, where the system can be non-dispersive. 

The on-site interaction between the atoms can be tuned using Feshbach resonances, which have become an important tool in ultracold atom collisions. The use of optical Feshbach resonances allows a faster tuning of the interactions in a spatially resolved way than magnetic fields, allowing for a sensitive control of the frustration. By cooling bosonic and fermionic atoms to ultracold temperatures\cite{Banerjee:2012pg} and trap them in an optical lattice formed by lasers we can simulate the chaotic-integrable transition, the non-Fermi liquid behaviour\cite{sachdevf} near saturation and study the nature of quantum chaos and its saturation bound in the SYK model. 

Kagome lattices have been studied for long time in condensed matter physics due to their unique properties associated with a large degree of geometric spin frustration, a phenomenon of having a large number of degenerate ground states for several geometric reasons, forbidding any type of ordering at zero temperature, and leading to exotic phases of matter at large $N$ limit. 

The realization of SYK model in real $d$-dimensions requires a momentum independence of the spectrum and therefore a flat band. For a single particle, the geometrical frustration leads to the appearance of a flat band associated to the quenched dispersion in the lattice\cite{Eberlein:2017wah}. A Kagome lattice, which consists of corner-sharing triangles, has a flat excited band via geometric frustration in the band structure. If the optical lattice is sparsely populated by the interacting particles, the kinetic energy of the system will be quenched and the complete state of the lattice only depends on the minimization of the interactions. 
Therefore, the trapped atoms in a flat band are strongly correlated and will not have any kinetic energy, while their dynamics will only be determined by their interactions and topology in the system. The kinetic energy is always quenched in the flat band and as the atom interactions become stronger, novel interaction-driven phases, such as the trion-superfluid or supersolid may appear. As the strength of fermion interaction increases, the particle spectrum becomes flatter in the vicinity of the Fermi energy, towards forming a plateau, due to an increase of the effective fermion mass at the Fermi level and the corresponding peak in the density of states. 

In experimental context, placing the optical lattice in a high magnetic field, leads to a flattening of the energy levels. Kagome optical lattices with strong disorder already exhibit a flat band, therefore the atoms don't have to be placed in a strong magnetic field. The flatness of the band is generated by the destructive interference with degenerate localized states between the hexagons of the Kagome structure. The lowest motional band is split into three sub-bands, while for a deep lattice, in the tight-binding limit, the uppermost sub-band will be flat. The normal velocity of the ultracold atoms in the optical lattice is proportional to the curvature of the band they occupy. The high degree of frustration of the Kagome geometry is reflected by the presence of non-dispersive orbital bands. 

The exotic SYK physics and the chaotic non-Fermi liquid behaviour can be concretely realized via the interplay of disorder and random interactions in the optical lattice. For the SYK model\cite{remarks}, a low energy effective theory of trapped spinless fermions with strong disorder emerges and therefore a consequent emergent conformal symmetry will be spontaneously broken, leading to zero modes and exhibiting a maximal Lyapunov exponent in the chaos region. SYK is maximally chaotic because the out-of-time order correlators exhibit Lyapunov exponents and a butterfly effect, saturating the chaos bound from black hole theory\cite{boundml}. 

\section{SYK model, chaos and black holes}

SYK model is a concrete solvable model with non-Fermi liquid behavior and maximal chaos, describing a collection of randomly interacting Majorana fermions with significant connections to black hole theory and quantized solutions of gravity. Its connections to dilaton gravity has recently attracted considerable attention from interdisciplinary community, especially high energy, field-theoretical and condensed matter physics research groups. It has been a subject of major interest in the last decade, especially on the gravity side of the theory, due to its remarkable properties (emergence of conformal symmetry in the IR limit, the effective action, the four-point functions and chaos), allowing the probing of non-equilibrium dynamics and scrambling of information associated with black holes. This enables the study of phase transitions, quantum critical points, and the robustness of emergent symmetries under varying conditions of disorder.

At the same time, the model gives rise to thermalization and many-body quantum chaos with a Lyapunov exponent that saturates the quantum limit, like a black hole in dilaton gravity. Specifically, SYK is closely related to 2d dilaton gravity, describing the excitations near the horizon of extremal black holes. Both connections to quantum chaos and information scrambling, therefore to the black hole information paradox make SYK a suitable toy model for quantum simulation of black holes. 
The appearance of emergent conformal symmetries at the critical infrared fixed point and a divergent contribution of the symmetry modes in the propagator of the bi-local field corresponding to the dilaton-gravity sector in the dual AdS theory are key properties of SYK model. Furthermore, it has also been conjectured that the SYK model may describe the low energy limit of a higher dimensional gauge theory with a string theory dual, where the coupling of SYK bulk states is similar to the discrete states of 2d string theory.

The strong interest in the SYK model is related to the low temperature characteristics which are similar to those of strong gravity conditions in the infrared (IR) limit described by AdS$_2$ geometry, such as a finite entropy at zero temperature, a ground state energy extensive in the number of particles, and a specific heat linear in temperature but with a prefactor different from free fermions. As SYK becomes strongly interacting at low energies, in this limit, its gravity-dual interpretation can be inferred in the low-temperature strong-coupling limit. While the duality is only present at large $N$ number of atoms and a low energy limit (conformal limit), the realization of the SYK nearly conformal behavior and its dual black hole in the nearly AdS$_2$ space is extremely useful to black hole theory in general. 

SYK model at large \( N \) is characterized as a ``fast scrambler'' redistributing localized excitations across the many-body degrees of freedom of a system. This behavior is closely related to quantum chaos, thermalization, and entanglement generation, occurring at the maximum possible rate. A common method for diagnosing scrambling is by observing the early-time exponential decay of out-of-time-order correlators (OTOCs), which is quantified by the quantum Lyapunov exponent \( \lambda_L \). In the strong coupling regime \( \beta J \gg 1 \), the SYK model saturates the universal bound \( \lambda_L \leq 2\pi/\beta \), representing the upper limit on the speed of quantum information scrambling. This property is shared with black holes, suggesting a holographic interpretation of the model. The connection is further supported by its effective (disorder-averaged) action. In the infrared (IR) limit, the effective action is conformally invariant and reproduces the large-\( N \) Schwinger--Dyson equations, which give rise to the Green’s function solutions. The specific form of the Green’s function spontaneously breaks the conformal symmetry down to \( \text{SL}(2,\mathbb{R}) \), with corrections to the IR effective action described by the Schwarzian action. These features are also characteristic of two-dimensional Jackiw–Teitelboim gravity.

SYK model has attracted attention due to its integrability in the large $N$ limit\cite{adscft11}, with approximate conformal symmetry in the IR limit. SYK exhibits properties characteristic of black holes, such as an extensive ground state entropy, an emergent conformal symmetry  at low energies and the fast scrambling of quantum information saturating the universal bound on the Lyapunov chaos exponent. The SYK flows to a conformal theory in deep IR, leading to the existence of a bulk dual of the theory. The model can be studied at large $N$ limit, leading to an emergent reparametrization invariance at the IR critical point, with OTOCs associated to quantum chaos. 
SYK exhibits nonlocal fermion interaction and the presence of non-Fermi liquid states with non-zero entropy density at vanishing temperature. At large $N$ limit, where the SYK model is solvable, the two-point correlation function exhibits its non-Fermi liquid behavior, while the OTOC function shows a maximal Lyapunov exponent. In $N$ limit, the SYK model acquires conformal symmetry and its effective action can be approximated by the Schwarzian action\cite{Stanford:2017thb}. The large-$N$ solution displays a time-reparametrization (conformal) invariance at low energies, leading to a connection to black holes and string theory.
The IR and near-IR limit are both solvable, with the invariant Schwarzian action representing the boundary gravitational degrees of freedom, associated with Jackiw–Teitelboim (JT) dual theory i.e. two-dimensional ``near-$AdS_2$'' gravity with dilaton coupling. The leading correction to the out-of-time ordered four-point correlation function grows exponentially with time, and the growth exponent saturates the chaos bound, a similar behaviour to the chaotic dynamics of black holes. 

The SYK theory behaves as a CFT at low energies, while the infrared limit of the model has an AdS$_2$ bulk gravity dual. A two-dimensional JT gravity becomes the holographic dual of the IR fixed point of the SYK model, via the AdS/CFT correspondence as a bridge connecting the conformal field theory (CFT) in $d$ dimensions and its gravity dual in the AdS background in $d+1$ dimensions. JT theory is effectively one-dimensional, as the dynamics is only dependent on the boundary curve of AdS spaceime. In the infrared limit, the JT effective action exactly coincides with the SYK effective action. Unfortunately, the JT model properties are only equivalent with the lowest-energy features of SYK model, given by the Schwarzian action and therefore don't define a complete gravity dual of SYK model.    
In gravity theory, in the dual AdS space, the properties of the black hole are embedded into the partition function.
If $Z$ is the partition function here, describing the correlations between all energy levels at all scales,
the correlations between neighbour energy levels are given by the distribution of $r_n$. The associated spectral form factor will be
\begin{equation}
\begin{aligned}
&Z(J/T+iJ\tau)Z(J/T-iJ\tau)
=&\sum_{n, m} e^{-(J/T+iJ\tau) E_{n}} e^{-(J/T-iJ\tau) E_{m}},
\end{aligned}
\end{equation}
By studying the spectral form factor, one can make statements about the Hilbert space of black holes. 
If $A$ is a local unitary operator and $dA$ is the Haar measurement associated with it, the averaged two-point OTOC function is $\int dA\left\langle A(0)A^{\dagger}(t)\right\rangle_{J}$ and it is proportional to the spectral form factor $|Z(2T,\tau)|^{2}$. 
A realization of the SYK Hamiltonian has a spectral form factor with large fluctuations at late time. In this context, if we take the ensemble average the spectral form factor, any fluctuations will smooth out.
The self energy becomes asymptotically exact in the limit $N\to \infty$ and has a solution in the low-frequency limit. Assuming time-translation invariance, the fermion propagator is associated with the self energy via the Dyson equation
\begin{equation}\label{syk3}
G(\omega_n)=[-i\omega_n-\Sigma(\omega_n)]^{-1}, 
\end{equation}
defined as
$G(\omega_n)=\int_0^\beta d\tau e^{i\omega_n\tau}G(\tau)$, where the Matsubara frequencies are $\omega_n=\pi T(2n+1)$
with $n$ integer.

From the self energy diagrams, the
bare propagator  $G_0(\omega_n)=-1/i\omega_n$ and the dressed
propagator $G(\omega_n)$ we determine the  propagator at large $N$
\begin{equation}\label{syk3b}
\Sigma(\tau)=J^2G^3(\tau),
\end{equation}
where the local fields $G, \Sigma$ have exact solutions to the mean field equations in large $N$ limit, 
ignoring the fluctuations around the saddle point values, as solutions of the Schwinger-Dyson
equations. At lowest energies, SYK model keeps its quantum characteristics by avoiding replica symmetry breaking.

We here consider the complex fermion SYK model (cSYK) model of a system of $N$ spinless fermions in (0+1) dimensions subject to random all-to-all fermion interactions, dual to dilaton gravity in (1+1) dimensional AdS$_2$ space. The cSYK model at half filling exhibits a holographic dual to a black hole. The SYK model with complex fermions considers a thermal state with the chemical potential and the mass term in the Hamiltonian turned on, where a first order phase transition forms. The high temperature phase shows chaotic behaviour (chaotic phase), however the low temperature phase is integrable (integrable phase).  The integrable phase can be described by a weakly interacting massive theory, where the Lyapunov exponent is near zero. Such phase transition shows similar properties to the Hawking-Page transition between the black hole phase and thermal AdS phase. The common key element is the equivalent partition function between the gravity theory and the SYK model.

The second-quantized Hamiltonian of spinless fermions $c_i$ is
\beq
H = \frac{1}{(2 N)^{3/2}} \sum_{i,j,k,\ell=1}^N J_{ij;k\ell} \, c_i^\dagger c_j^\dagger c_k^{\vphantom \dagger} c_\ell^{\vphantom \dagger} 
- \mu \sum_{i} c_i^\dagger c_i^{\vphantom \dagger}, \label{H}
\eeq
with the canonical fermions obeying
\beq
c_i c_j + c_j c_i = 0 \quad, \quad c_i^{\vphantom \dagger} c_j^\dagger + c_j^\dagger c_i^{\vphantom \dagger} = \delta_{ij}, \label{anti}
\eeq
and the $J_{ij;k\ell}$ are complex random variables, independent Gaussian couplings with zero mean obeying
\bea
J_{ji;k\ell} = - J_{ij;k\ell} \quad , \quad
J_{ij;\ell k} &=& - J_{ij;k\ell} \quad , \quad
J_{k\ell;ij} = J_{ij;k\ell}^\ast \nn
\overline{|J_{ij;k\ell}|^2} &=& J^2.
\eea
There is only a fermion interaction term in $H$, and no fermion hopping, working as a `matrix model'
on the Fock space, with a dimension exponentially large in $N$.
In order to realize the SYK Hamiltonian, $J_{ij;k\ell}$ must be close to random independent variables with a
Gaussian distribution.

The density $0<\mathcal{Q}<1$ is U(1)conserved and it 
depends on a a fermion number constraint, quantified as the average fermion number
\beq
\mathcal{Q} = \frac{1}{N} \sum_i \left\langle c_i^\dagger c_i^{\vphantom \dagger} \right\rangle, \label{defQ}
\eeq
on every site $i$ and $0 < \mathcal{Q} < 1$ that can be controlled by the chemical potential $\mu$. The exchange interactions $J_{ij;k\ell}$ are independent Gaussian random numbers with zero mean and equal variance.
As the interactions between the fermions are all-to-all and completely random, there will be no
distance between fermions and therefore the Hamiltonian will be zero-dimensional. 

\begin{figure}[h!]
    \centering
    \includegraphics[width=0.99\textwidth]{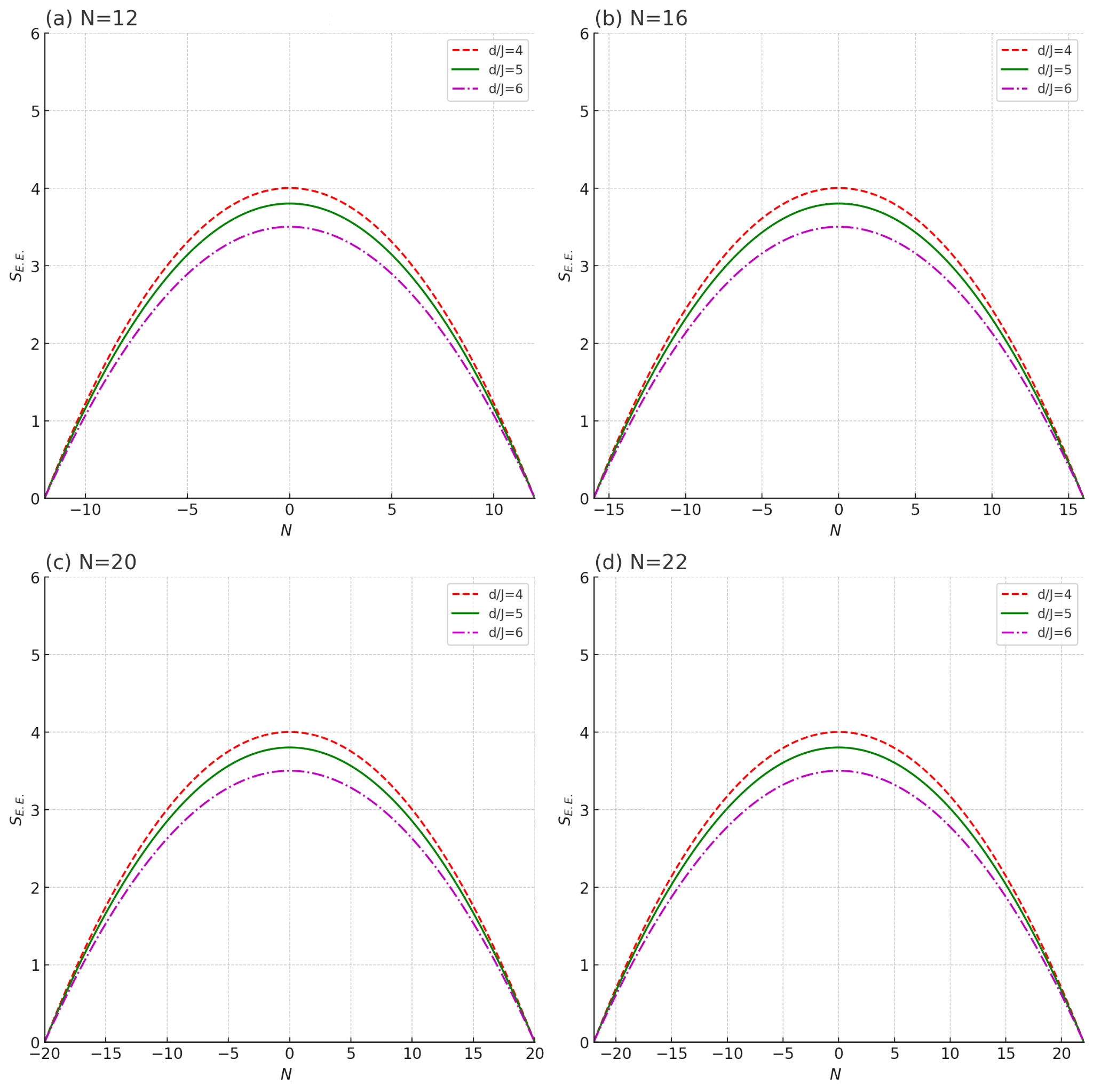}
    \caption{Ground-state entanglement entropy for different choices of discrete distance $d/J$ and for different lattice system sizes $N_A$=}
    \label{fig:EE}
\end{figure}

The ground state of the complex-SYK model is characterized by bipartite entanglement that follows a volume law. Figure~\ref{fig:EE} shows the entanglement entropy as a function of the lattice separation d. The entanglement entropy $S_{E.E.}$ is calculated using the von Neumann entropy of the reduced density matrix of a subsystem $A$ with $N_A$ consecutive sites:

\begin{equation}
    S_{E.E.} = -\text{Tr}(\rho_A \ln \rho_A),
\end{equation}

\begin{equation}
    \rho_A = \text{Tr}_{\bar{A}}(|G\rangle \langle G|),
\end{equation}
where $A$ and $\bar{A}$ represent a partition of the lattice’s spatial degrees of freedom,  $\rho_A$ is the reduced density matrix and $|G\rangle$ is the ground state of the Hamiltonian.

The fermions are integrated out and the action is solved in the saddle-point approximation in large $N$ limit, with the
saddle-point equations
\bea 
P_{ab} (\tau, \tau') &=&  \left\langle c_{a}^\dagger (\tau) c_{ b}^{\vphantom\dagger} (\tau') \right\rangle \nn
Q_{ab} (\tau, \tau') &=&  J^2 \left| P_{ab} (\tau, \tau') \right|^2
\eea
defined as a single-site problem, as all sites are identical. The diagonal solutions are
\beq
P_{ab} (\tau, \tau') = \delta_{ab} G(\tau' - \tau),
\eeq
with $G(\tau)$ the fermion Green's function defined as
\beq
G(\tau, \tau') = -\frac{1}{N} \sum_i \left\langle T_\tau \left( c_i (\tau) c_i^\dagger (\tau') \right) \right\rangle
\eeq
with $T_\tau$ denotes Euclidean time-ordering, and $G(\tau-\tau') = G(\tau, \tau')$.
The large $N$ saddle-point equations is
\beq
G(i \omega_n) = \frac{1}{i \omega_n + \mu - \Sigma (i\omega_n)} \quad, \quad \Sigma (\tau) = -  J^2 G^2 (\tau) G(-\tau), \label{eq:SY}
\eeq
with $\omega_n$ as Matsubara frequency.  

The strong connection between the SYK model and quantum chaos is quantified by the amount of chaos in a quantum system (scrambling), as a process in which the quantum information locally deposited in the system, will be distributed among all degrees of freedom. Black holes scramble with maximum possible efficiency, by saturating the fundamental bound on the Lyapunov exponent. At low temperatures, the emergent conformal symmetry and chaos described by the Schwarzian mode associated to the breaking of the conformal symmetry can be made equivalent to the boundary graviton field of the dilaton-gravity in $2d$. The gaps in the SYK spectrum are exponentially suppressed in $N$. The quantum chaos dynamics near the black hole event horizon, in the context of gauge/gravity duality is given by the behavior of OTOCs and the exponential growth of the corresponding commutators.

A generalized statement of the AdS/CFT correspondence asserts that the black hole solution corresponds to the thermal ensemble in the boundary quantum theory, and the quantum thermalization is dual to the black hole formation in the bulk gravity theory. In the bulk gravity context, black holes are fast scramblers and their chaotic behaviour leads to scrambling, parametrized by the Lyapunov exponent with an upper bound saturated by the black hole solution. The out of time ordered four point correlators in SYK model saturates the upper bound on the Lyapunov exponent. 
In a quantum quench, the parameters of the Hamiltonian are abruptly changed, starting from an equilibrium configuration, such as the ground state of the system.  The change in the coupling excites the system that will evolve with a final Hamiltonian. 
The Lyapunov exponent has a finite value in the chaotic phase and it is effectively zero in the integrable phase.
An interesting problem is how an OTOC grows in the fermion lattice and how the bound saturates to a black hole equivalent solution. The influence of an initial perturbation can be quantified by the size of a commutator $[V(0),W(t)]$, where the operator $V(0)$ is the initial perturbation and $W(t)$ and
quantifies its influence at the time $t$. The OTOC $C(t)={\rm tr}(e^{-\beta H} [V(0),W(t)]^2)$ at the time $t$ can be
calculated. As the OTOC is given by the thermal average of THE Loschmidt echo signals, OTOCs can be measured via the many-body Loschmidt echo method, based on the idea that small perturbations of the Hamiltonian may trigger dramatic changes of the dynamics, inducing the butterfly effect.

\begin{figure}[h!]
    \centering
    \includegraphics[width=0.99\textwidth]{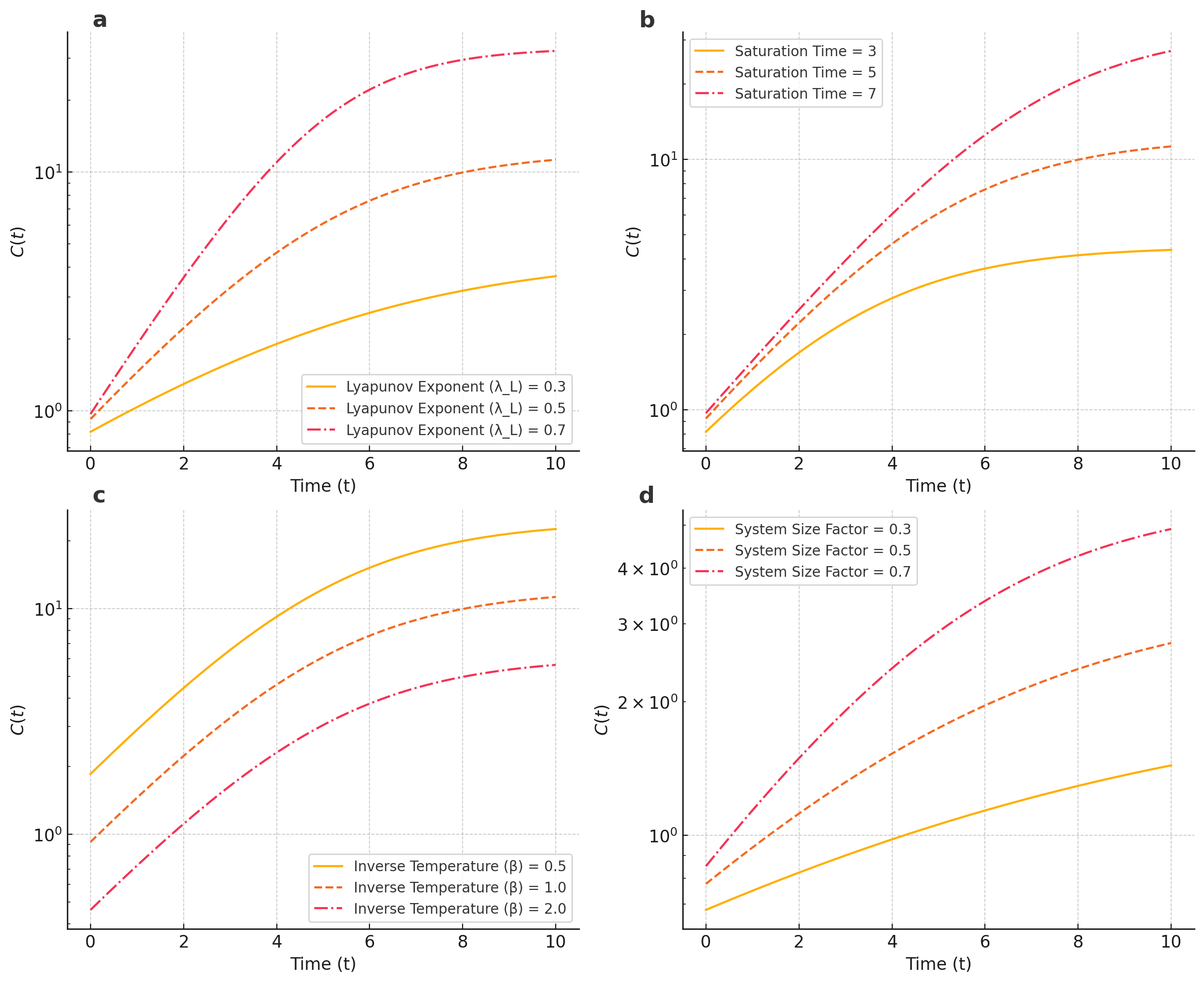}
    \caption{(a) OTOC growth for different values of the Lyapunov exponent ($\lambda_L$), illustrating the effect of chaotic strength on the system's evolution.
   (b) OTOC growth for various saturation times, representing systems with different interaction timescales, showing how the point of saturation shifts.
    (c) Temperature effects on OTOC growth, with different values of inverse temperature ($\beta$), demonstrating how quantum scrambling changes at varying temperatures.
    (d) System size effects, using different scaling factors, highlighting how the OTOC growth and saturation change with system size variations.}

    \label{fig:OTOCs}
\end{figure}

The Figure~\ref{fig:OTOCs} illustrates the dependence of OTOC dynamics on the Lyapunov exponent ($\lambda_L$), saturation time, inverse temperature ($\beta$), and system size, showing that larger $\lambda_L$ values lead to faster exponential growth, longer saturation times slow information scrambling, lower temperatures suppress chaos, and larger system sizes exhibit delayed OTOC saturation, reflecting how quantum chaos and the phase transition from chaotic to integrable phases are governed by these parameters in the discrete-real-SYK model.

If $V,W$ are initially commuting Hermitian operators, the operator $C(t)$ will increase exponentially, due to
the scrambling of information when the perturbation spreads throughout the system. Experimentally, the averaged two-point OTOC function can be measured at large $N$ and low energy, leading to the study of chaotic behavior with the bound saturated.
The quantum Lyapunov exponent $\lambda_L$ represents the exponent of the exponential growth.
The conformal structure generates the OTOC $C(t)$ with a characteristic exponential growth 
at the maximal permissible Lyapunov exponent $\lambda_L = \frac{2
  \pi}{\beta}$, holographically connecting the model to a black hole.

The complex fermion SYK model physics is implemented on the lattice by a general Hamiltonian containing two species of fermions
$\mathcal{H}=\mathcal{H}_{K}+\mathcal{H}_{imp}+\mathcal{H}_{int}$, containing a tight-binding Hamiltonian $\mathcal{H}_{K}$ of cold fermionic atoms with a flat band loaded in the optical lattice (due to the fine tuning of the chemical potential), an impurity Hamiltonian $\mathcal{H}_{imp}$ given by a number of randomly distributed $\delta$-function potentials, and a short-range particle interaction Hamiltonian $\mathcal{H}_{int}$. The role of the impurity Hamiltonian $\mathcal{H}_{imp}$ is to stabilize the localization in the lattice but still relating the wave-functions between nearest-neighbour cells. By removing some states from the flat band, $\mathcal{H}_{imp}$ breaks the rotational symmetry of the lattice, resulting in a smeared spectrum. 

The fermionic atoms are confined and controlled in potential wells of the optical Kagome lattice, where
each well has a number of strongly localized atomic states filled with a number of atoms.
The associated tight-binding Hamiltonian on the lattice $\mathcal{H}_{K}$ has a general form:

\begin{equation}\label{tight}
\begin{split}
    \mathcal{H}_{K} = & -\mu\sum_{m}\left(a_{\mathbf{r}_{m}}^{(a)\dagger}a_{\mathbf{r}_{m}}^{(a)}
    + a_{\mathbf{r}_{m}}^{(b)\dagger}a_{\mathbf{r}_{m}}^{(b)} 
    + a_{\mathbf{r}_{m}}^{(c)\dagger}a_{\mathbf{r}_{m}}^{(c)}\right) \\
    & - t\sum_{\langle m,n \rangle} e^{i\varphi} \left( 
    a_{\mathbf{r}_m}^{(b)\dagger}a_{\mathbf{r}_n}^{(a)} 
    + a_{\mathbf{r}_m}^{(a)\dagger}a_{\mathbf{r}_n}^{(c)} 
    + a_{\mathbf{r}_m}^{(c)\dagger}a_{\mathbf{r}_n}^{(b)}
    \right) + h.c.
\end{split}
\end{equation}
with $\mu$ the fermion chemical potential keeping the Fermi surface in the flat band, $t$ the hopping term, $a_{\mathbf{r}_{m}}^{(\alpha)\dagger}$ and $a_{\mathbf{r}_{m}}^{(\alpha)}$($\alpha=a,b,c$) the creation and annihilation fermion operators localized at $\mathbf{r}_{m}$ on a sub-lattice $\alpha$. Here, the phase of the hopping $\varphi$ is introduced to help tuning the gauge fields\cite{Wei:2020ryt}. 

A number of randomly distributed impurities in the lattice with a potential $V$, will generate the impurity Hamiltonian 
\begin{equation}\label{impurity}
\mathcal{H}_{imp}=V\sum_{{r}_{m}\in R}a_{\mathbf{r}_{m}}^{\dagger}a_{\mathbf{r}_{m}},
\end{equation}
where $R$ is a random set of $M$ sites in the lattice.
If $A(\mathbf{r})$ is a short-range two-body interaction, the corresponding interaction Hamiltonian is
\begin{equation}\label{interaction}
\mathcal{H}_{int}=\frac{1}{2}\sum_{mn}\l_{\mathbf{r}_{m}}A(\mathbf{r}_{m}-\mathbf{r}_{n})\l_{\mathbf{r}_{n}},
\end{equation}
where $\l_{\mathbf{r}_{m}}=a_{\mathbf{r}_{m}}^{\dagger}a_{\mathbf{r}_{m}}$ is the particle number operator on each site $m$.

\begin{figure}[h!]
    \centering
    \includegraphics[width=0.8\textwidth]{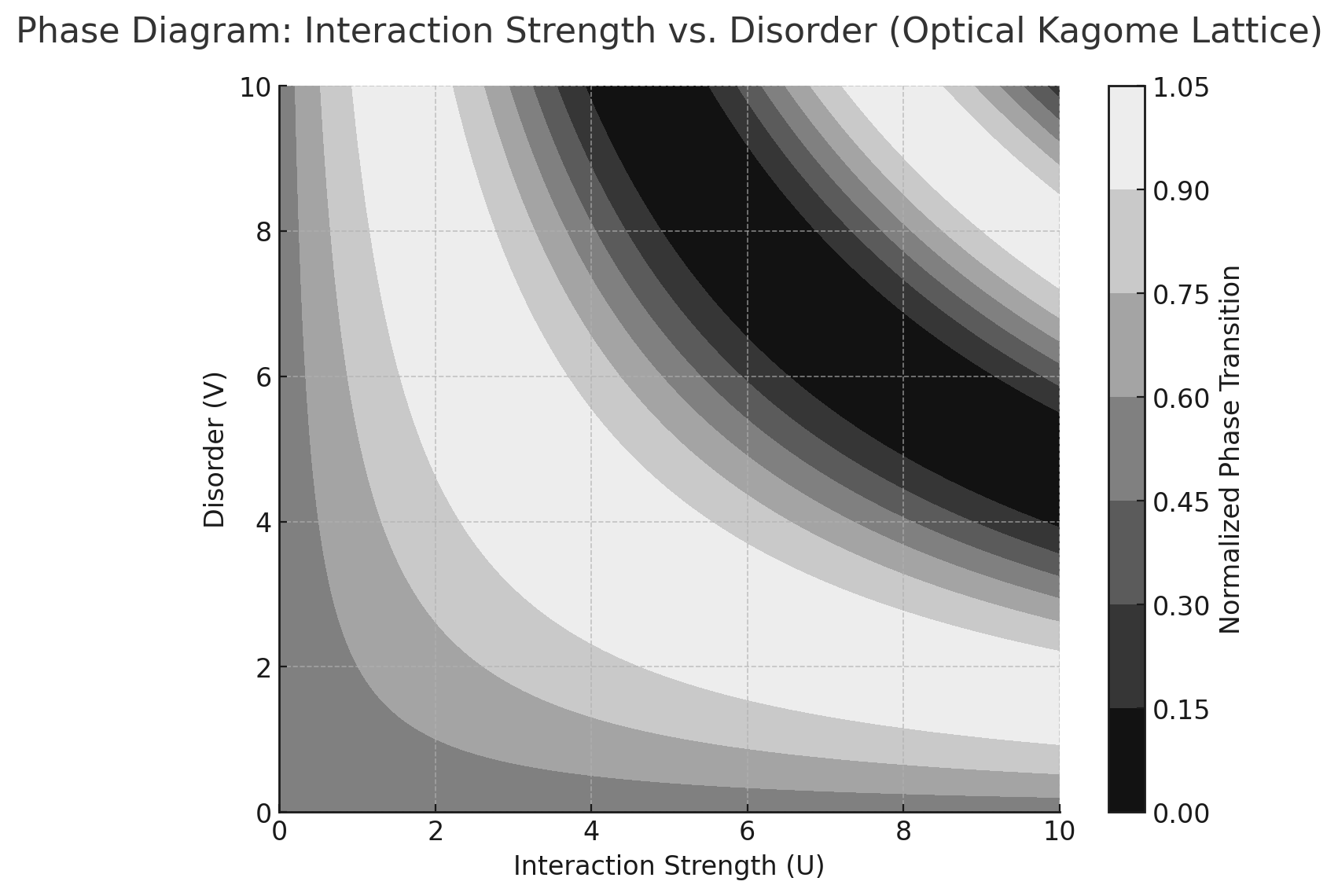}
    \caption{Phase Diagram: Interaction Strength vs. Disorder (Optical Kagome lattice), based on the effective Hamiltonian in Eq. 17. The plot visualizes how the system's quantum phases evolve as interaction strength (U) and disorder (V) are varied, for a fixed hopping term. The system transitions between different phases, with the lighter regions representing higher interaction or disorder, likely corresponding to more chaotic or non-Fermi liquid behavior.}
    \label{fig:phase_diagram}
\end{figure}

Figure~\ref{fig:phase_diagram} illustrates how different phases emerge on the Kagome lattice depending on the interaction strength (tuned via Feshbach resonances) and the level of disorder (randomness in atomic placement). The chaotic phase occurs when both parameters are high, while integrable phases occur when both are low, with crossover regions in between. The contour plot visualizes how different phases of the system (e.g., chaotic and integrable phases) transition based on these two variables. The phase transition occurs as both interaction strength and disorder increase, with the system becoming more chaotic and less integrable at higher values of both parameters. The darker regions in the diagram correspond to phases with lower interaction strength and disorder, representing more integrable, less chaotic behavior. Lighter regions, on the other hand, represent highly disordered, strongly interacting phases, where the SYK model's properties—such as conformal symmetry breaking, maximal Lyapunov exponents, and non-trivial quantum critical behavior—are likely to manifest. In the Kagome lattice with strong disorder, a flat energy band forms due to geometric frustration, where the kinetic energy is quenched. As interaction strength increases, the system moves toward a non-dispersive regime, mimicking the non-Fermi liquid behavior of the SYK model. The phase diagram reflects this flattening of the spectrum as interaction strength and disorder increase, resulting in phases dominated by interaction-driven phenomena, rather than kinetic energy. 

For $N$ particles with the wave-functions $\phi_{i}(\mathbf{r}_{m}), i=1,\cdots, N$, if the interaction is sufficiently weak and the Fermi surface lies in the flat band, the SYK physics is generated by the free Hamiltonian $\mathcal{H}_{K}+\mathcal{H}_{imp}$ and the perturbation $\mathcal{H}_{int}$ independent of the nature of the short-range two-particle interaction $\mathcal{H}_{int}$. The second quantized wave function of the fermion at site ${\bf r}_m$ takes a general form $a_{\mathbf{r}_{m}}=\sum_{i} \phi_{i}({\bf r}_m) c_i$ over the basis of the flat-band wave functions, with $c_i$ as annihilation operator. 

Intriguingly, placing the atoms in a Kagome lattice adds a low-energy orbital degree of freedom, caused by placing the particle on each site within a unit cell and thus generating non-trivial ordering and dynamics.
The low-temperature effective Hamiltonian for the degenerate ground states is
\begin{equation}
\mathcal{H}_{\text{eff}}=(F_{\varphi}-\mu)\sum_{i}c_{i}^{\dagger}c_{i}+\sum_{ijkl}\tilde{J}_{ijkl}c_{i}^{\dagger}c_{j}^{\dagger}c_{k}c_{l},
\end{equation}
where the random four fermion coupling $\tilde{J}_{ijkl}$ takes the form
\begin{equation}
\tilde{J}_{ijkl}=\frac{1}{2}\sum_{\mathbf{r}_{1}\mathbf{r}_{2}}[\phi_{i}(\mathbf{r}_{1})\phi_{j}(\mathbf{r}_{2})]^{*}V(\mathbf{r}_{1}-\mathbf{r}_{2})[\phi_{k}(\mathbf{r}_{1})\phi_{l}(\mathbf{r}_{2})],
\end{equation}
with $\phi_{i}(r)$ the wave function of the $i$-th degenerate state, $r_{1/2}$ the lattice sites and $c_{i}^{\dagger}$ and $c_{i}$ are creation and annihilation fermion operators. Using the anti-commutation relations of creation and annihilation operators, the effective Hamiltonian
$\mathcal{H}_{\text{eff}}$ becomes
\begin{equation}\label{eff}
\mathcal{H}_{\text{eff}}=(F_{\varphi}-\mu)\sum_{i}c_{i}^{\dagger}c_{i}+\sum_{i>j,k>l}J_{ijkl}c_{i}^{\dagger}c_{j}^{\dagger}c_{k}c_{l},
\end{equation}
with
\begin{equation}\label{J}
J_{ijkl}=\tilde{J}_{ijkl}+\tilde{J}_{jilk}-\tilde{J}_{jikl}-\tilde{J}_{ijlk}.
\end{equation}
where $F_{\varphi}$ is a tuning parameter that depends on the phase of the hopping $\varphi$ to control the artificial gauge field\cite{Aidelsburger:2018rfk}. 
Experimentally, the strength of the atomic interaction and the phase of the hopping can be controlled by Feshbach resonances, occurring when two atoms in well-defined spin states collide and couple to a virtual state with a different spin configuration, inducing a coupling between the lowest energy bands for strongly confined atoms.

\section{Conclusions}

Besides simulating quantum systems and quantum computing, optical lattices are a promising tool for studying black holes, holographic principle and AdS/CFT conjecture. The model provides a unique opportunity to study the effects of disorder and randomness in a controlled setting. Characteristic non-trivial signatures of the cSYK physics, such as the exponential growth of OTOCs, the solvability in the large $N$ limit, the emergence of conformal symmetry in IR and saturation of the bound on many-body quantum
chaos, can be realized via a randomly interacting ultracold atoms, optically trapped in a Kagome lattice. The random couplings inherent to SYK model, which are analogous to those found in black hole horizons, can be implemented on an optical lattice with controlled imperfections.

We show that the SYK model as the low energy effective theory of spinless fermions (using ultracold atoms with nontrivial properties) placed in an optical Kagome lattice with strong disorder is a viable toy model to study the gravity dual of 2-dimensional black holes in a tractable fermionic Hamiltonian and generate maximally chaotic behaviour similar to chaos in the near-AdS$_2$ spacetimes. By coupling the system to another system of peripheral fermions defined by the impurity Hamiltonian, interesting  physics from a phase transition into a Fermi liquid phase emerges, inducing new properties and behaviour characteristic to black holes in AdS spacetime.
The combination between strong interactions, controlled disorder, and tunable artificial gauge fields in the optical lattice enables the study of chaos and correlated transport phenomena in SYK model. 

Most importantly, the proposed lattice setup offers a potential SYK toy model for studying black holes together with the connection between quantum black holes and interacting quantum many-body systems in the near future, via the AdS/CFT correspondence. The SYK model’s capacity to exhibit maximal chaos, characterized by a Lyapunov exponent that saturates the bound predicted for quantum chaotic systems, provides a good framework to understand the rapid scrambling of information in black hole dynamics, providing valuable data for refining models of quantum gravity. 

In the large-$N$ and strong coupling limit, SYK has characteristic properties resembling a black hole, such as the same entropy density as a black hole in AdS, several similar correlation functions and a Lyapunov exponent pattern for quantum theories with dual gravity description. 
The characteristic emergent conformal invariance of SYK model at low energies, its extensive ground state entropy, and the OTOC dynamics define the paradigmatic relationship between the SYK model and black holes. \cite{Sachdev-1006} 


\end{document}